\documentclass[aps,prb,onecolumn,showpacs,12pt]{revtex4-1}
\usepackage{graphicx}
\usepackage{color} 
\usepackage{amssymb}
\usepackage{natbib}
\usepackage{hyperref}
\usepackage{amsmath}

\begin{document}

\preprint{APS/123-QED}

\title{Effect of pressure on the physical properties of the superconductor NiBi$_3$}

\author{Elena Gati$^{1,2}$, Li Xiang$^{1,2}$, Lin-Lin Wang$^{1}$, Soham Manni$^{1,2}$, Paul C. Canfield$^{1,2}$, and Sergey L. Bud'ko$^{1,2}$}

\address{$^{1}$ Ames Laboratory, Iowa State University, Ames, Iowa 50011, USA, and\\$^{2}$ Department of Physics and Astronomy, Iowa State University, Ames, Iowa 50011, USA}

\date{\today}

\begin{abstract}
We present an experimental study of the superconducting properties of NiBi$_3$ as a function of pressure by means of resistivity and magnetization measurements and combine our results with DFT calculations of the band structure under pressure. We find a moderate suppression of the critical temperature $T_c$ from $\approx\,4.1\,$K to $\approx\,3\,$K by pressures up to 2\,GPa. By taking into account the change of the band structure as a function of pressure, we argue that the decrease in $T_c$ is consistent with conventional, electron-phonon-mediated BCS-type superconductivity.
\end{abstract}

\maketitle

%\maketitle

\section{Introduction}

Bismuth-rich compounds have played a leading role in the last couple of years in the search for novel topological phases \cite{Isaeva13} due to the strong spin-orbit coupling which is associated with the heavy Bi atoms. Among these systems, Bi$_2$Se$_3$ and Bi$_2$Te$_3$ were established as good candidate systems for topological insulators \cite{Xia09,Chen09}. The concept of topology, however, is not only restricted to insulators, but can also be extended to any system with a gap in the excitation spectrum \cite{Shao16}. This includes, in particular, superconductors which, in case they are topological in nature, have potential applications in quantum computing \cite{Nayak08}. It is argued, that unconventional $p$-wave superconductors intrinsically realize this novel topological state \cite{Sato17}. Among the Bi-based compounds, Cu$_x$Bi$_2$Se$_3$ \cite{Sasaki11} is a prominent example to realize such a topological superconducting state in the Bi-based family.

Another Bi-rich compound, which is known to superconduct, is NiBi$_3$. It has a critical temperature $T_c$ of $\approx\,4\,$K \citep{Alekseevskii52}. However, the nature of superconductivity in this compound has not yet been clarified. In studies of the nature of superconductivity (SC), pressure has proven to be an useful tool to explore the properties of SC (see e.g. \cite{Lorenz05} for a review). In the case of NiBi$_3$, a very early report suggested an increase of the superconducting critical temperature $T_c$ with pressure $p$ \cite{Alekseevskii52}. However, a detailed study of the properties of NiBi$_3$ under pressure is still missing. Thus, we present here a comprehensive study of the superconducting properties of NiBi$_3$ under pressure by utilizing measurements of the resistance and magnetization, combined with DFT calculations of the band structure under pressure. Our results show a decrease of $T_c$ with $p$ which can be described quantitatively within the framework of BCS theory. Thus, our results indicate a conventional nature of superconductivity in NiBi$_3$. 

%interplay of ferromagnetism and superconductivity; Bi-based compounds topological stuff; spin-orbit coupling; earlier report of increasing Tc with p

\section{Experimental and Computational Details}

Single crystals of NiBi$_3$ were grown by a flux-growth technique \cite{Canfield92} out of excess Bi. To this end, Ni (Alfa Aesar 99.99\%) and Bi  (Alfa Aesar 99.99\%) in the molar ratio 1:9 were loaded into a fritted alumina Canfield Crucible Set \cite{Canfield16} and sealed into a fused silica ampoule under partial argon atmosphere. The ampoule was heated to 1000$^\circ$C in 3 hours and dwelled at this temperature for another 3 hours. It was then cooled rapidly to 600$^\circ$C (3 hours) and dwelled there for another 3 hours. After that, the ampoule was slowly cooled to 320$^\circ$C over 40 hours. At this temperature, the ampoule was removed from the furnace and excess liquid was decanted by the help of a centrifuge. The resulting crystals were needle-shaped and had dimensions of about $\approx\,5\,\times\,0.2\,\times\,0.2$\,mm$^3$ (see bottom inset of Fig.\,\ref{fig:fig13}). These crystals were characterized initially by x-ray as well as magnetization measurements. The x-ray data was acquired by using a Rigaku MiniFlex diffractometer (Cu $K_{\alpha1,2}$ radiation) at room temperature on ground single crystals.

Magnetization at ambient pressure was measured in a Quantum Design Magnetic Property Measurement System (MPMS-3). An aggregate of randomly oriented single crystals of total mass $m\,\approx\,24\,$mg were loaded in a gelantine capsule. The background of the gelantine capsule and the sample holder was determined independently and subtracted from the measured magnetization data.\\
For measurements of the resistivity under pressure, the samples were cut into pieces of dimensions of $\approx\,2\,\times\,0.2\,\times\,0.2$\,mm$^3$. Resistivity was measured in a standard four-point configuration; four 25\,$\mu$m Pt wires were attached to the sample by using silver epoxy. The ac resistivity measurements were performed in a Quantum Design Physical Property Measurement System with a current of 1\,mA and frequency of 17\,Hz. The current was applied along the long needle axis which is the crystallographic $b$ axis. Measurements were performed upon cooling using a rate of $-$\,0.25\,K/min. For measurements in magnetic field, $\mu_0 H$, the field was applied perpendicular to the long needle axis, i.e., perpendicular to the crystallographic $b$ axis. To apply pressures up to 2\,GPa, the sample was placed in a Be-Cu/Ni-Cr-Al hybrid piston-cylinder cell. The design of the cell is similar to the one described in Ref. \cite{Budko84}. A 4:6 mixture of light mineral oil:n-pentane \cite{Budko84,Thompson84,Kim11} was used as a pressure-transmitting medium. It solidifes at $\sim\,$3$-$4 GPa at room temperature, i.e.,  well above our maximum pressure. Thus, hydrostatic pressure conditions are ensured in the full pressure range of investigation. Pressure at low temperatures was inferred from the pressure dependence of the superconducting temperature $T_c$ of Pb \cite{Bireckoven88}. \\
Magnetization measurements under pressure were performed in MPMS-3 in a dc field of 2\,mT. A commercially-available HDM Be-Cu piston-pressure cell \cite{Mcell} was used to apply pressure up to 1.2\,GPa. Daphne oil 7373 was used as a pressure medium \cite{Yokogawa07}, which solidifies at 2.2 GPa\,at room temperature, thus ensuring hydrostatic pressure conditions in the full $p$ range. $T_c(p)$ of Pb was used to determine the low-temperature pressure \cite{Bireckoven88}.

The band structure of NiBi$_3$, at different hydrostatic pressures, was calculated in density functional theory \cite{Hohenberg64,Kohn65} (DFT) using PBEsol as exchange-correlation functional with the spin-orbit coupling (SOC) effect included. With the forces and stress calculated using the Feynman-Hellmann theorem, the unit cell lattice constants and ionic positions were fully relaxed at different pressures before the band structures were calculated. The density of states (DOS) was calculated using the tetrahedral method in reciprocal space. All DFT calculations were performed in VASP \cite{Kresse96a,Kresse96b} with a plane-wave basis set and projector augmented wave \cite{Bloechl94} method. We used the orthorhombic cell of 16 atoms with a $\Gamma$-centered Monkhorst-Pack (5$\,\times\,$12$\,\times\,$4) k-point mesh \cite{Monkhorst76} and kinetic energy cutoff was 337 eV. The convergence with respect to k-point mesh was carefully checked, with total energy converged below 1 meV/atom. For ionic relaxation, the absolute magnitude of force on each atom was reduced below 0.01 eV/\AA. 

\section{Results}

	\subsection{Experimental Results}
	
		\subsubsection{Ambient-pressure characterization}
		
\begin{figure}
	\centering
	\includegraphics[width=0.8\textwidth]{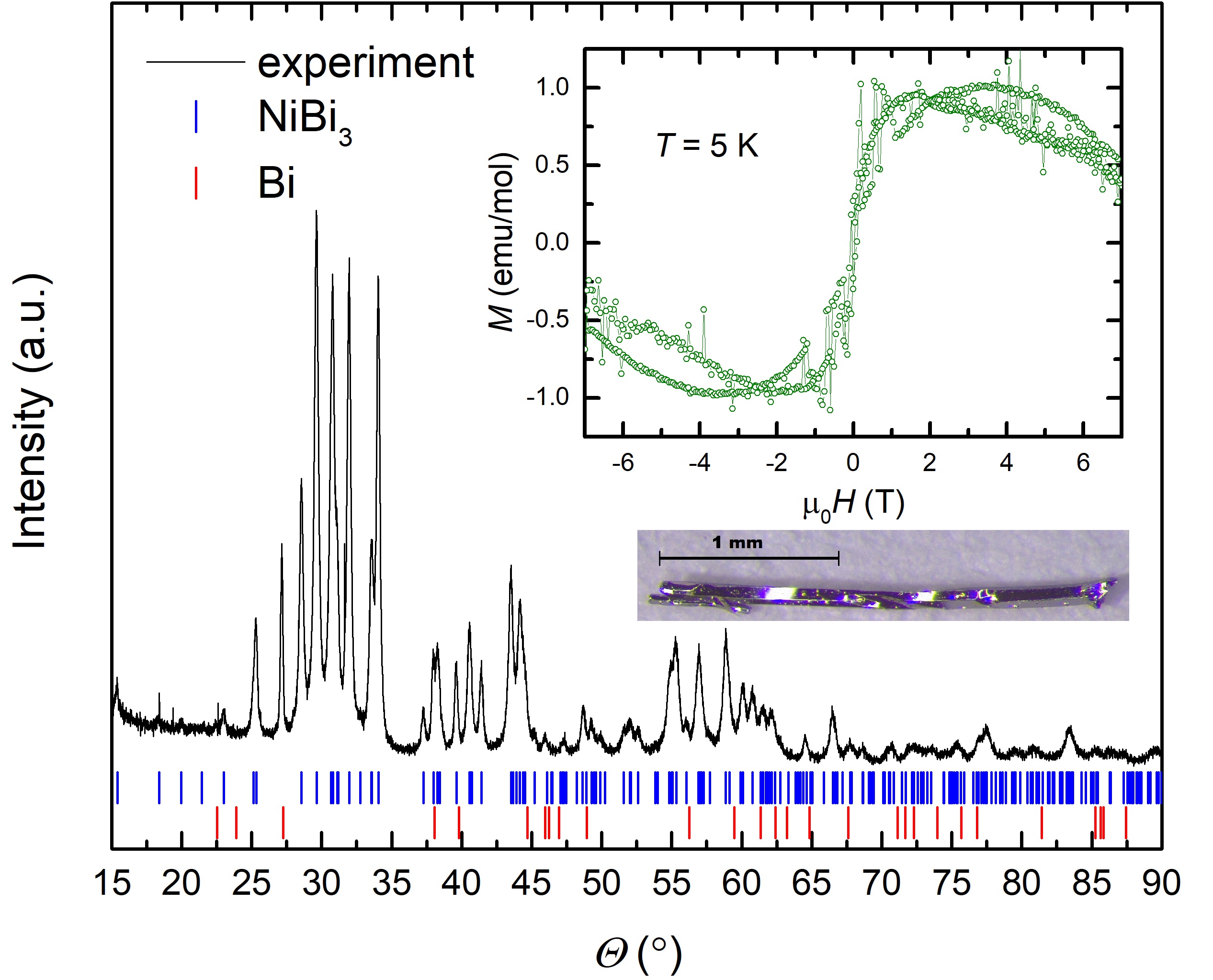} 
	\caption{Powder x-ray diffraction pattern (black line) and reported peak position of NiBi$_3$ (blue) and Bi (red); Top inset: Magnetization $M$ vs. field $\mu_0 H$ at constant temperature $T\,=\,5\,$K at ambient pressure; Bottom inset: Picture of a NiBi$_3$ crystal.}
	\label{fig:fig13}
	\end{figure}

		Figure \ref{fig:fig13} shows our ambient-pressure x-ray as well as magnetization data (inset). The peaks resolved in our powder x-ray diffraction pattern match very well with the reported peak positions of the orthorhombic $Pnma$ structure of NiBi$_3$. We also find additional peaks that can be attributed to excess Bi flux.

	In light of a previous report of amorphous Ni in flux-grown single crystalline NiBi$_3$ \cite{Silva13}, we note that our x-ray pattern does not indicate the presence of Ni impurities. However, as peaks from Ni might not be resolvable in x-ray diffraction due to small amount of Ni below the resolution limit and/or the amorphous nature of Ni impurities, we also performed measurements of the magnetization as a function of field, $\mu_0 H$, at $T\,=\,$5\,K, i.e., at $T\,>\,T_c$ (see Fig. \ref{fig:fig13}, top inset). This magnetization data set indicates weak ferromagnetism in the measured lump of NiBi$_3$ samples. The measured moment is very small and corresponds to $\approx\,10^{-4}\,\mu_B$ per formula unit.  As our data is comparable to the literature results presented in Ref. \cite{Silva13}, we infer that small amounts of Ni impurities are present in our samples, most likely in droplets of excess Bi flux.

		\subsubsection{Pressure-dependent measurements}
	
\begin{figure}
	\centering
	\includegraphics[width=0.6\textwidth]{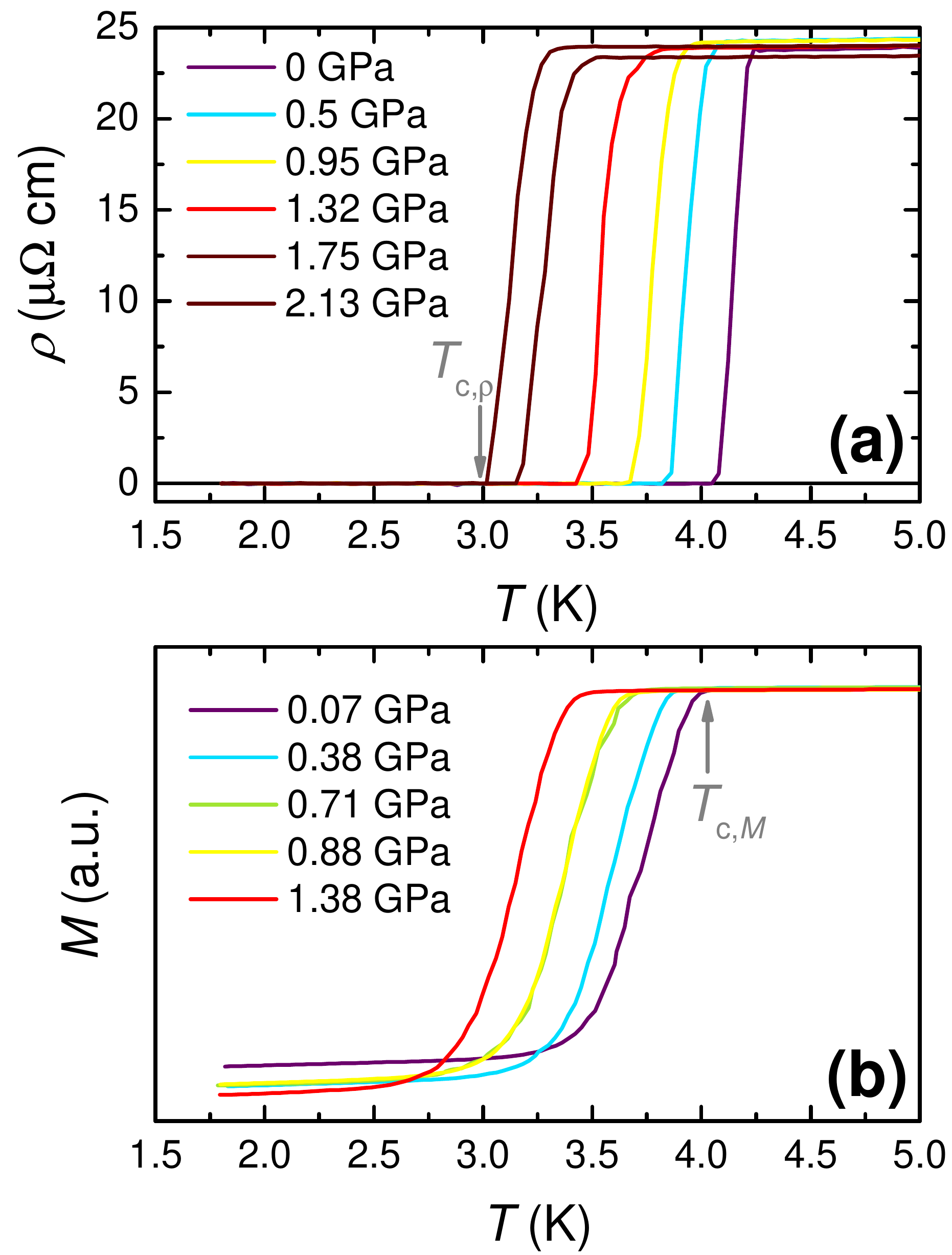} 
	\caption{Resistivity $\rho(T)$ (a) (sample \#1) and magnetization $M(T)$ (b) of NiBi$_3$ (sample \#2) as a function of temperature $T$ at different pressures $p$ up to 2.13\,GPa (a) and 1.38\,GPa (b). Resistance data was taken upon cooling. Magnetization data were taken upon warming. The criteria to determine the critical temperature from resistivity $T_{c,\rho}$ and magnetization $T_{c,M}$ are marked exemplarily by grey arrows for the data taken at $p\,=\,2.13\,$GPa (a) and 0.07 GPa (b).}
	\label{fig:fig1}
	\end{figure}

	First, we focus on a discussion of the effect of pressure on the superconducting properties of NiBi$_3$ and discuss the change of the critical temperature $T_c$ with pressure $p$. Figure \ref{fig:fig1} presents resistivity, $\rho$, and magnetization, $M$, data as a function of $T$ for various pressures up to 2.1\,GPa and 1.2\,GPa, respectively. At ambient pressure, we find a relatively sharp resistance transition at $T\,\approx\,4.1$\,K corresponding to the onset of superconductivity. The sharp feature in resistance is accompanied by a clear drop in $M(T)$ which reflects the onset of diamagnetism. This $T_c$ is very consistent with various reports in literature on samples grown with different methods, which typically should give rise to different ferromagnetic impurity concentrations. This rules out a significant influence of the Ni impurities on the superconducting properties of NiBi$_3$.  Upon applying pressure, the transitions seen in $\rho(T)$ and $M(T)$ shift to lower temperatures. Importantly, we do not find any significant broadening of the features in $\rho$ or in $M$ even up to the highest pressures of our experiment. This is consistent with the good hydrostatic conditions provided by our pressure environment.
	
\begin{figure}
	\centering
	\includegraphics[width=0.6\textwidth]{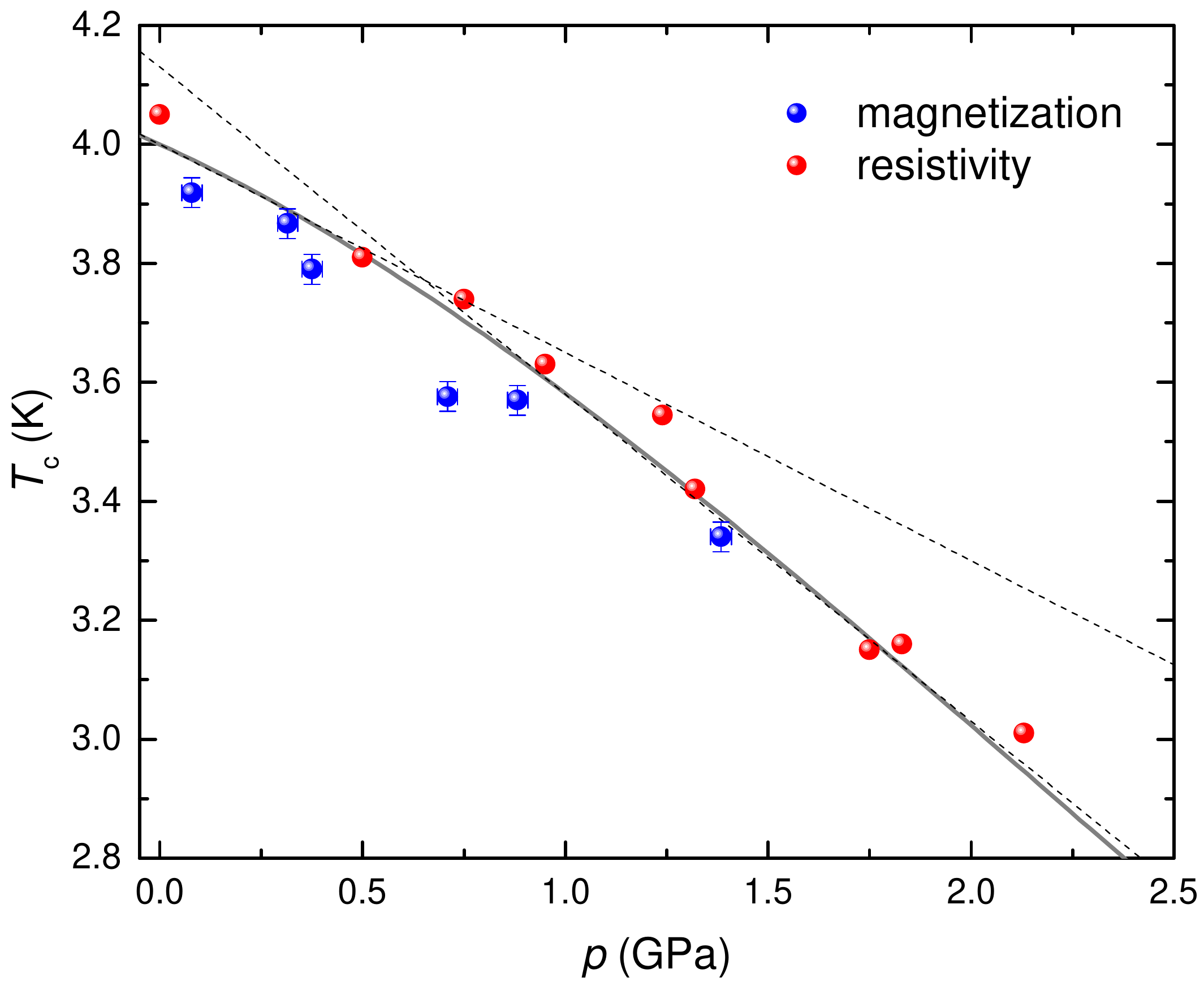} 
	\caption{Pressure, $p$, dependence of the superconducting transition temperature $T_c$ of NiBi$_3$  as inferred from magnetization (blue circles) and resistivity measurements (red circles). The slightly smaller $T_c$ values, inferred from magnetization  measurements, are most likely related to the shift of $T_c$ by the application of a small magnetic field, inevitable for measurements of the dc magnetization. Grey line is a guide to the eye. Dashed lines visualize the initial slope d$T_c/$d$p\,=\,-\,0.35$\,K/GPa and the slope d$T_c/$d$p\,=\,-\,0.55$\,K/GPa at $p\,=\,2$\,GPa.}
	\label{fig:fig2}
	\end{figure}

	To compile a temperature-pressure phase diagram from the data presented in Fig.\,\ref{fig:fig1}, we use the following criteria to determine $T_c$ from the present data sets (see arrows in Fig.\,\ref{fig:fig1}): The critical temperature from resistivity $T_{c,\rho}$ is defined as the temperature at which the resistance reaches zero. The critical temperature from magnetization $T_{c,M}$ is inferred from the onset of the drop in $M(T)$, i.e., the temperature at which the magnetization reaches 5\,\% of the value at 1.8\,K. The resulting $T$-$p$ phase diagram is presented in Fig.\,\ref{fig:fig2}. It should be noted that we find a somewhat lower value of $T_{c,M}$ compared to $T_{c,R}$ for all pressures investigated. This can most likely be related to the shift of $T_c$ by the small field of 2\,mT applied in measurements of the magnetization. Although, the extracted ambient-pressure values are in good agreement with previously-published literature results \cite{Alekseevskii52,Fujimori00}, we find an unambiguous decrease in $T_c$ with increasing pressure $p$. This is in contrast to earlier results \cite{Alekseevskii52}. Initially, $T_c$ is suppressed with a rate of d$T_c$/d$p\,\simeq\,-0.35$\,K/GPa. Upon increasing the pressure, $|$d$T_c$/d$p|$ nearly doubles and reaches a value of up to d$T_c$/d$p\,\simeq\,$-0.55\,K/GPa at $p\,=\,$2\,GPa (see dashed lines).
	
\begin{figure}
	\centering
	\includegraphics[width=0.8\textwidth]{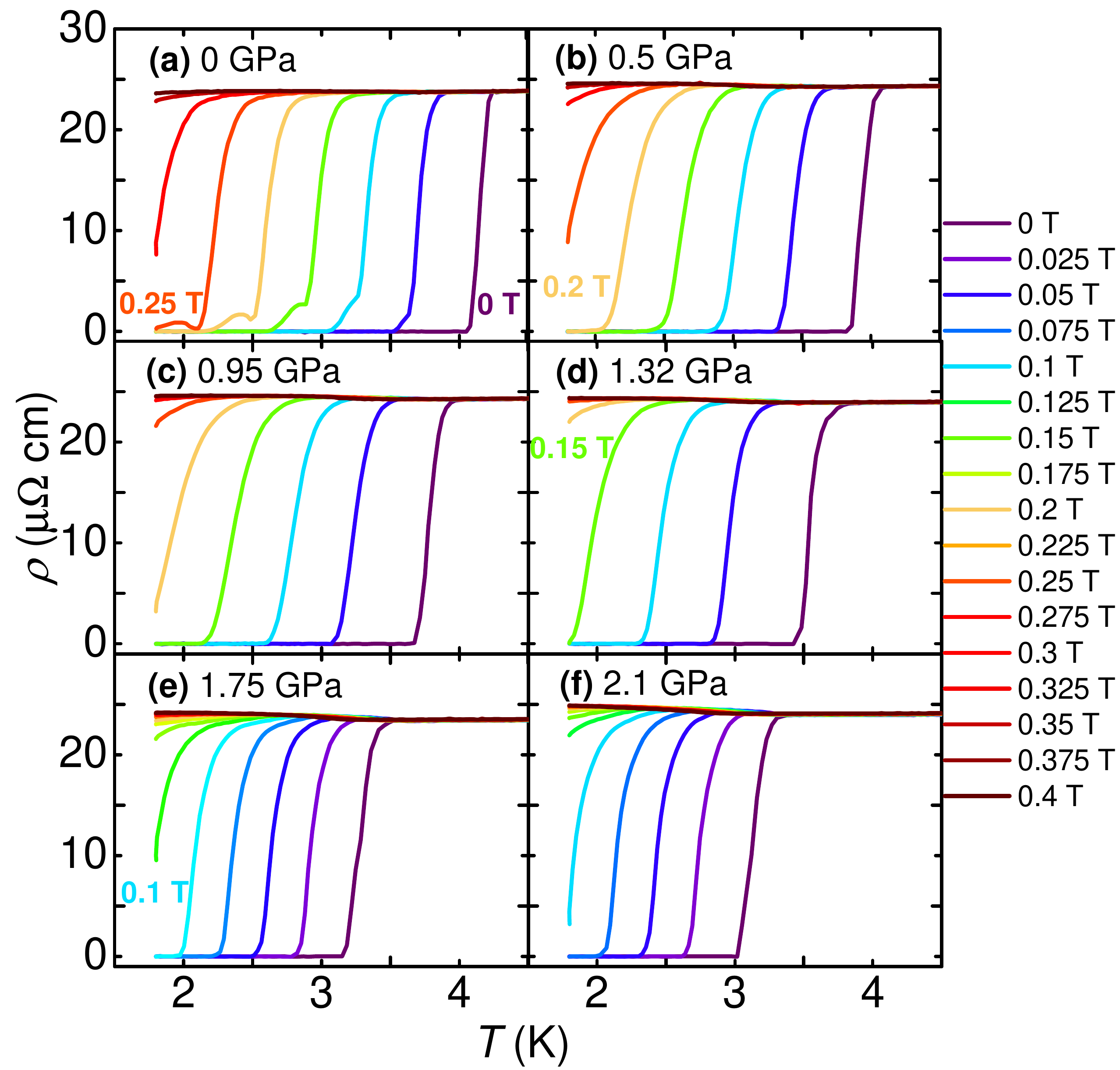} 
	\caption{Resistitvity $\rho(T)$ of NiBi$_3$ (sample \#1) at different constant fields $\mu_0 H\,\le\,0.4\,$T below $T\,=\,4.5\,$K at pressures ranging from 0\,GPa (a) to 2.1\,GPa (f). Field was applied perpendicular to the long needle axis ($H\,\perp\,b$ axis).}
	\label{fig:fig3}
	\end{figure}

\begin{figure}
	\centering
	\includegraphics[width=0.6\textwidth]{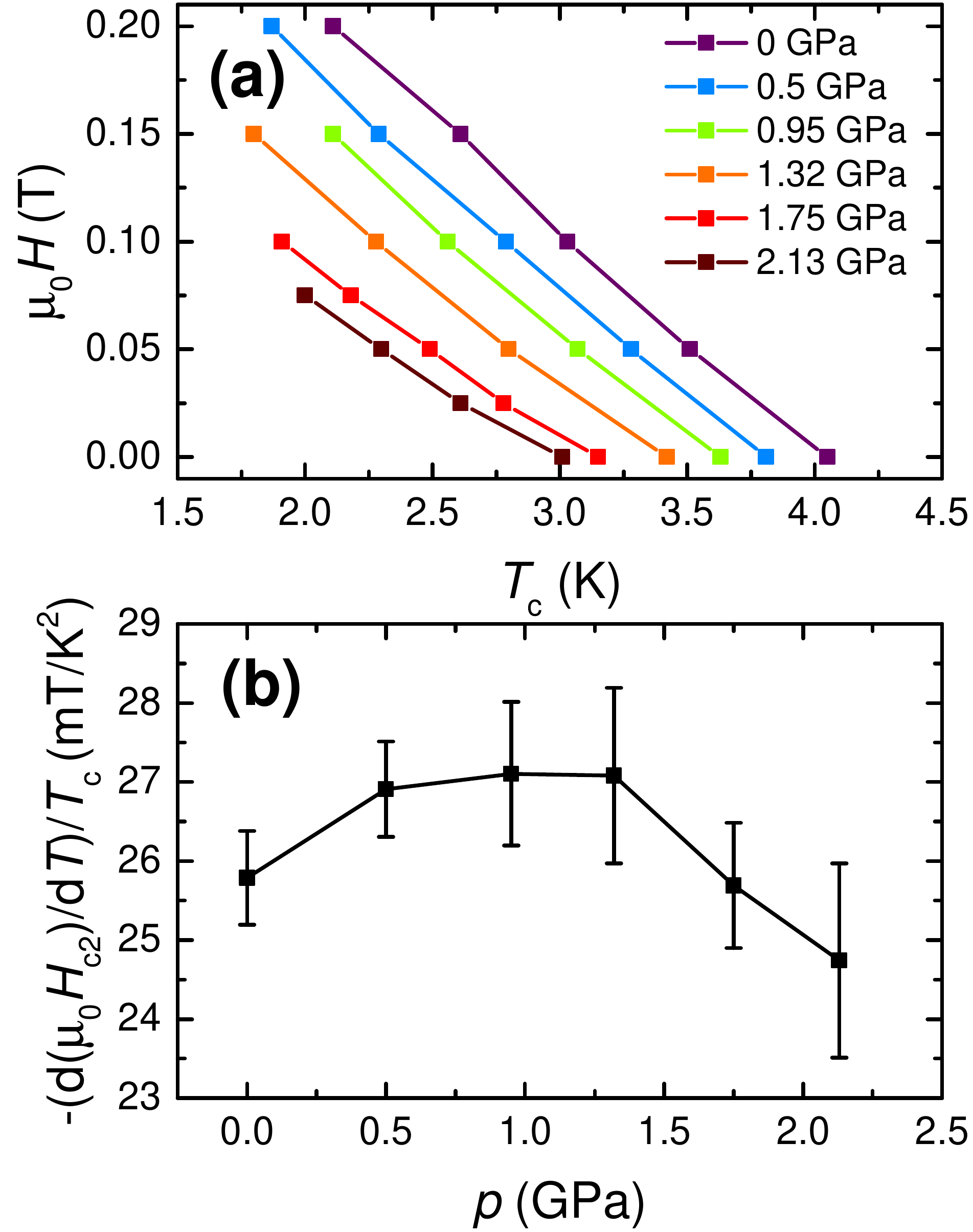} 
	\caption{(a) Field, $\mu_0 H$, dependence of the superconducting critical temperature $T_c$ of NiBi$_3$ (sample \#1) at different constant pressures; (b) Pressure dependence of the normalized slope of the upper critical field $H_{c2}$, -(d($\mu_0 H_{c2}$)/d$T$)/$T_c$, which is related to the pressure dependence of the Fermi velocity $v_F$ (see main text for details).}
	\label{fig:fig4}
	\end{figure}

	To provide a further characterization of the superconducting state under pressure, we studied the response of the superconducting transition to external fields, $\mu_0 H$, to obtain the pressure dependence of the upper critical field $H_{c2}$. Figure\,\ref{fig:fig3} shows the behavior of $\rho(T)$ at different fields between 0\,T and 0.4\,T at six representative applied pressure values. For all measurements shown, the field was applied perpendicular to the long needle axis of the crystal, i.e., perpendicular to the crystallographic $b$ axis. At ambient pressure, relatively small fields suppress $T_c$ until the zero-resistance state cannot be reached in a field of 0.25\,T down to 1.8\,K, the lowest temperature for these measurements. We stress that upon applying field the resistance feature remains fairly sharp. Only in a small field range 0.125\,T$\,\le\,\mu_0 H\,\le\,0.25$\,T a small kink at the low-temperature side of the jump is observed. The origin of the kink is unknown at present. However, at higher pressures this kink is absent and we find only a small increase in broadening in small fields. Our data (see Fig.\,\ref{fig:fig4} (a)) show that $H_{c2}$ is reduced (along with $T_c$) with increasing $p$: A field of 0.1\,T is sufficient at the highest pressure of 2.1\,GPa to not observe zero resistance at $T\,\ge\,1.8\,$K.
	
	A quantitative analysis of the pressure dependence of $H_{c2}$ is obtained by evaluating $T_c$ at different $H$ using the same criterion as defined above. The result of this analysis is presented in Fig.\,\ref{fig:fig4}\,(a). For all pressures, we find an almost linear dependence of $T_c$ with $H$ over the investigated temperature range 0.5$\,\lesssim\,T/T_c\,\lesssim\,$1. At ambient pressure the slope -(d$H_{c2}$/d$T$)$_{T_c}\,\approx$\,105\,mT/K. This linear $T$ dependence with similar slope (-(d$H_{c2}$/d$T$)$_{T_c}\,\approx\,158$\,mT/K for field applied in the same direction as in the present study) was already reported in a previous ambient-pressure study of the superconducting properties of NiBi$_3$ \cite{Fujimori00}. Upon increasing the pressure, the slope -(d$H_{c2}$/d$T$)$_{T_c}$ is reduced by $\approx\,$25$\%$ at $p\,\approx\,2\,$GPa.
	
	In general, the slope of $H_{c2}$ normalized by $T_c$, i.e., -(d$H_{c2}$/d$T$)$_{T_c}$/$T_c$, reveals information about the Fermi velocity, $v_F$, and the superconducting gap structure \cite{Kogan12,Taufour14}. In a single-band model, in the clean limit, it was shown that
	
	\begin{equation}
	-(\textnormal{d}H_{c2}/\textnormal{d}T)_{T_c}/T_c\,\propto\,\frac{1}{v_F^2}\ \cite{Kogan12}.
	\label{eq:slope}
	\end{equation}
	
	Even if such a single-band model likely represents a drastic simplification of the real band structure of the present material (see below for a detailed discussion of the band structure), changes in the normalized $H_{c2}$ slope might be attributed to changes of the Fermi surface and/or the superconducting gap structure. This formalism was already successfully applied in various multi-band pnictide superconductors, such as KFe$_2$As$_2$ \cite{Taufour14}, where abrupt Fermi surface changes at some critical pressure $p^*$ are reported in literature, or FeSe \cite{Kaluarachchi16}. However, the normalized slope of $H_{c2}$ as a function of $p$ in the present case of NiBi$_3$, presented in Fig.\,\ref{fig:fig4}\,(b), reveals only a smooth and broad maximum without any clear signatures of an anomalous behavior. This small change of $-(\textnormal{d}H_{c2}/\textnormal{d}T)_{T_c}/T_c$ with $p$ might indicate minor changes of the Fermi surface under $p$. These, in turn, might also be associated with the slight non-linearity in $T_c$ vs. $p$, discussed earlier. Nevertheless, due to the absence of a sudden change in $-(\textnormal{d}H_{c2}/\textnormal{d}T)_{T_c}/T_c$ we infer that no significant, abrupt change of the Fermi surface and the superconducting gap structure occurs over the investigated $p$ range.
	
\begin{figure}
	\centering
	\includegraphics[width=0.8\textwidth]{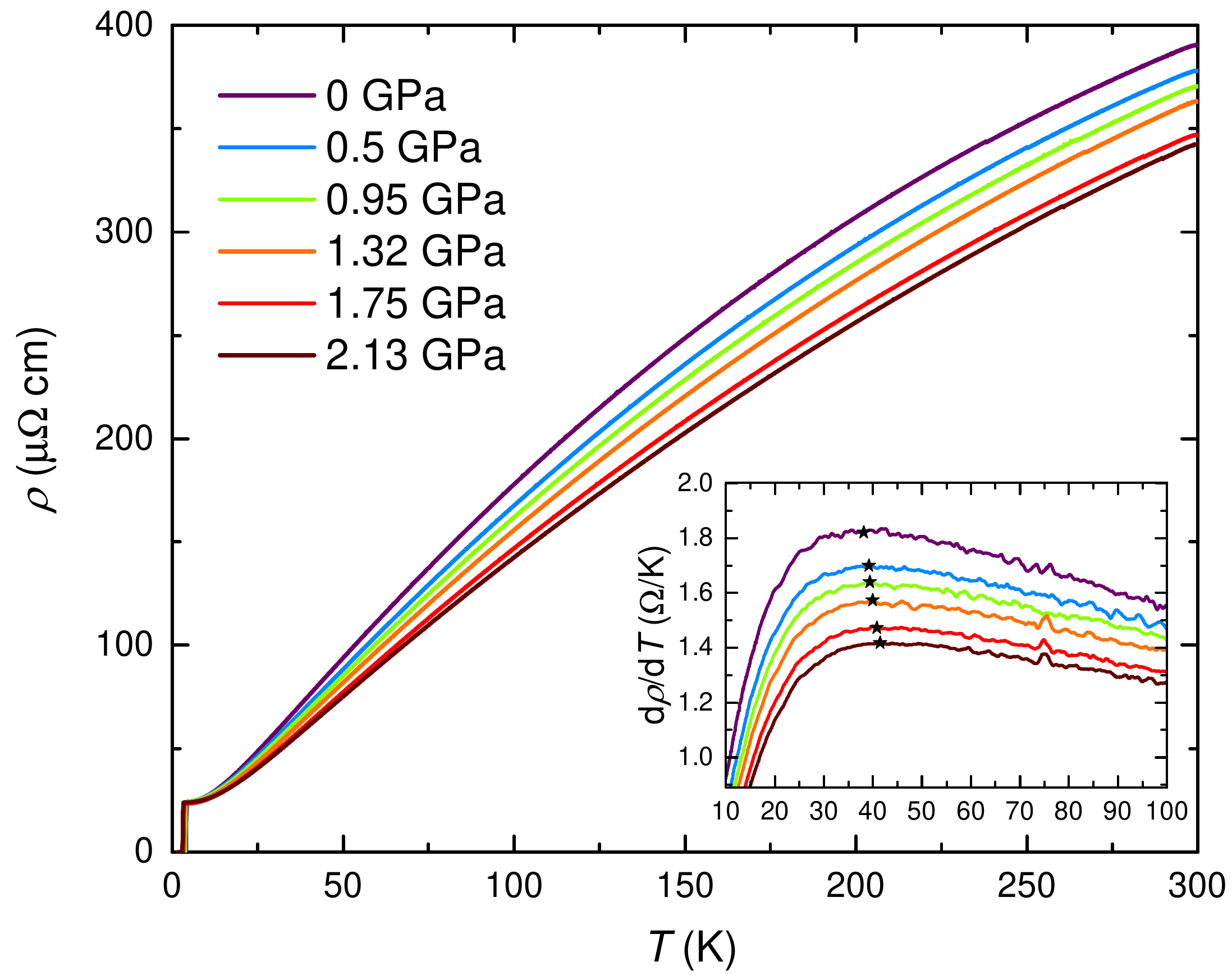} 
	\caption{Temperature-dependent resistivity, $\rho(T)$, of NiBi$_3$ (sample \#1) up to 300\,K at different applied pressures displaying the pressure dependence of the normal-state resistivity. All curves were taken upon cooling; Inset: Evolution of the maxima in d$\rho$/d$T$ with pressure. The change in the position of these maxima can be used to estimate the change of Debye temperature as a function of pressure (for details, see main text).}
	\label{fig:fig5}
	\end{figure}

	Next, we discuss our results on the normal-state properties of NiBi$_3$, based on a study of the resistivity up to room temperature. Figure  \ref{fig:fig5}\, shows $\rho(T)$ at different constant pressures for 1.8\,K$\,\le\,T\,\le\,300$\,K. Our ambient-pressure results are in agreement with previous literature results; we find a steep increase of $\rho$ above $T_c$ without indications for an extended $T$ regime in which $\rho\,\propto\,T^n$ holds. At $T\,\approx\,40$\,K, d$\rho$/d$T$ exhibits its maximum (see inset of Fig.\,\ref{fig:fig5}). At even higher temperatures $\rho$ increases monotonically, however, d$\rho$/d$T$ is reduced below 1 upon increasing $T$. Thus, $\rho$ tends toward saturation at high temperature ($T\,>\,300$\,K). The residual resistivity ratio ($RRR$\,=\,$\rho(300\,\textnormal{K})/\rho(4.3\,\textnormal{K})$) for the present sample is $\approx\,16.4$ and is thus as large as previously reported RRRs \cite{Fujimori00,Nedellec85} indicating a similar good quality of the single crystal used in the present study. Upon applying pressure, we find an overall reduction of $\rho(T)$ at high temperatures whereas the low-temperature resistance is almost unaffected. However, the overall behavior of $\rho(T)$ stays essentially the same. This includes the broad maximum in d$\rho$/d$T$ occuring at lower $T$ and the resistance tending to saturate at higher $T$. In particular, there is no evidence for any phase transition other than superconductivity occurring over investigated $p$-$T$ range. 
	
	Earlier studies \cite{Fujimori00,Nedellec85} of the resistivity behavior of NiBi$_3$ already pointed out the unusual $T$-dependence of $\rho$, i.e., the saturation of $\rho$ at high $T$, which deviates from the $\rho\,\propto\,T$ behavior accounted for by the Bloch-Gr\"uneisen model. The latter describes the $T$-dependent contribution of electron-phonon scattering to $\rho$. In previous reports, this discrepancy was explained by a reduction of the inelastic mean-free path at high $T$. However, the origin of this reduction is not yet clarified. Nevertheless, the resistivity behavior at intermediate temperatures ($10\,K\,<\,T\,\ll\,150\,$K) is likely to be dominated by electron-phonon scattering. The contribution of electron-phonon-scattering, as described by the Gr\"uneisen-Bloch model, typically gives rise to a maximum in d$\rho$/d$T$ at $T_{max}$. It indicates the crossover from the low-temperature $\rho \propto T^5$ for $T\,\ll\,\Theta_D$ behavior to the in-$T$ linear behavior dominating at high $T\,\gg\,\Theta_D$. The size of $T_{max}$ is directly related to the size of the Debye temperature $\Theta_D$, i.e., as larger $T_{max}$ the larger $\Theta_D$ is. Thus, the evolution of $T_{max}$ with pressure allows one to track the evolution of $\Theta_D$ with pressure. We find that $T_{max}$ increases from $\,\approx\,38.2\,$K at ambient pressure to $\,\approx\,41.5\,$K at $p\,\approx\,2.1\,$GPa. This corresponds to a d$\Theta_{max}$/d$p\,\approx\,+$1.5\,K/GPa. Thus, we infer an increase of $\Theta_D$ with $p$ which is typical for various materials (see e.g. \cite{Kaluarachchi17}) as the lattice usually hardens upon pressurization.

	\subsection{Theoretical DFT results}
	\label{sec:DFTresults}

	\begin{figure}
	\centering
	\includegraphics[width=\textwidth]{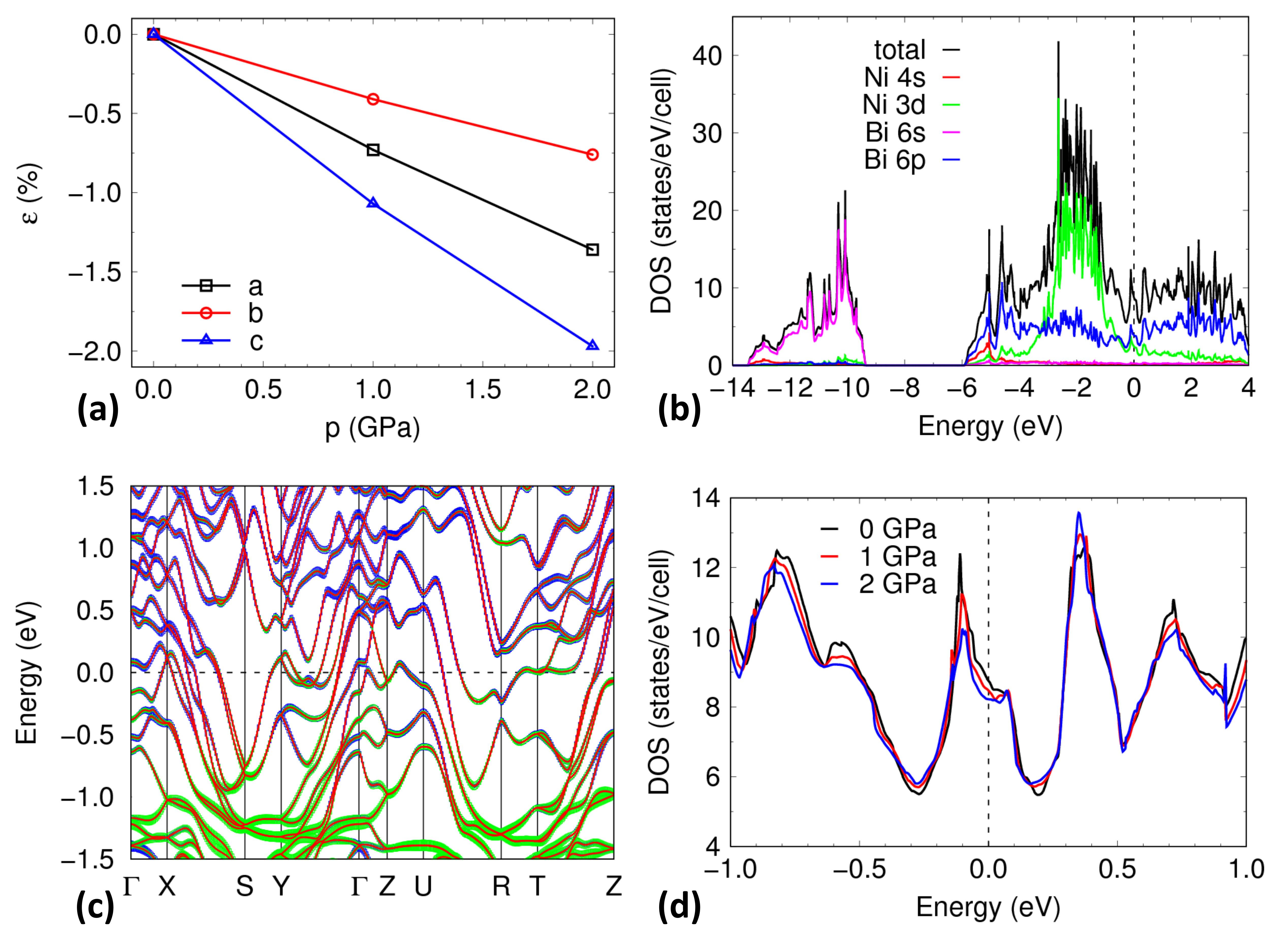} 
	\caption{(a) Strains, $\epsilon$, of NiBi$_3$ along the three crystallographic directions $a$, $b$ and $c$ under hydrostatic pressure $p$ of 1 and 2 GPa calculated with PBEsol+SOC in DFT; (b) Total and projected density of states (DOS) on Ni 4s, Ni 3d, Bi 6s and Bi 6p orbitals at $p$\,=\,0 GPa with the Fermi energy $E_F\,$=\,0 eV; (c) Band structures of NiBi$_3$ at $p\,=\,0$ GPa (orange lines) with green (blue) vertical bars stand for the relative magnitude of projections on Ni 3d (Bi 6p) orbitals; (d) Total DOS near $E_F$ with $p\,=\,$0, 1 and 2 GPa.}
	\label{fig:fig11}
	\end{figure}	
	
To study the effects of hydrostatic pressure on the properties of NiBi$_3$, we have carried out DFT band structure calculations at different pressures. The SOC is included in the DFT calculations to account for the presence of heavy Bi. We chose PBEsol as the exchange-correlation functional because it reduces the overestimation of bond length, a well-known problem for regular PBE. For NiBi$_3$ at zero pressure, the fully relaxed lattice constants with PBEsol+SOC are $a\,=\,8.861\,$\AA, $b\,=\,4.098\,$\AA\,and $c\,=\,11.388\,$\AA, which agree very well (within $\approx\,$1\%) with the experimental values \cite{Ruck06} of 8.879\,\AA, 4.0998\,\AA\ and 11.483\,\AA, respectively. The calculated bulk modulus is 38.5\,GPa. Figure \ref{fig:fig11}(a) shows the relative change of the lattice constants with respect to zero pressure, or strain $\epsilon\,=\,x(p)/x(p\,=\,0\,\textnormal{GPa})-1$ in each direction $x\,=\,a,b,c$, under hydostratic pressures of $p\,=\,1$ and 2 GPa. The strains are almost linear functions of pressure. At $p\,=\,$2 GPa, the strain along $c$ is the largest ($\epsilon_c\,=\,$−1.97\%), followed by the strain along $a$ ($\epsilon_a\,=\,$−1.36\%) and the strain along $b$ is the smallest ($\epsilon_b\,=\,$−0.76\%). This results in a 4\% reduction of the unit cell volume at $p\,=\,2$\,GPa, compared to the ambient-pressure unit cell volume.

Figure \ref{fig:fig11}\,(b) plots the total DOS at zero applied pressure as well as the DOS projected on Ni 4s, Ni 3d, Bi 6s and Bi 6p orbitals with the Fermi energy ($E_F$) set at 0 eV. The bands in the low energy range of $-$14 to $-$9 eV are derived mostly from Bi 6s orbitals and are separated by a gap from rest of the bands. The bands derived from Ni 4s orbitals are mainly in the energy range of $-$6 to $-$5 eV. Most of the features in the broad energy range from $-$6 to +4 eV are derived from Ni 3d and Bi 6p orbitals, which have strong band overlap and hybridization, agreeing with the earlier study \cite{Kumar11}. The main features of Ni 3d bands are located in the energy range from $-$3 to $-$1 eV and are almost filled. In a distinct contrast, the Bi 6p bands spread out in the full energy range. Right at the $E_F$, the contribution from Bi is 61\%, substantially larger than that from Ni 39\%. This result is different from an earlier study \cite{Kumar11}, where the projections on different Bi sites were not summed up. To confirm these orbital projection features, the band structure $E(k)$ together with orbital projections around $E_F$ is shown in Fig.\,\ref{fig:fig11}\,(c). There are 6 bands crossing the $E_F$ as clearly seen in the X-S direction. The green (blue) vertical bars on the band structure give the relative magnitude of projection on the Ni 3d (Bi 6p) orbitals. At lower energy, the bands dominantly have Ni 3d orbital character. Moving toward $E_F$, the orbital character of Bi 6p increases, while that of Ni 3d decreases. At the $E_F$, the bands have more orbital character from Bi 6p than Ni 3d, agreeing with the projected DOS results in Fig.\,\ref{fig:fig11}\,(b). 

Figure \ref{fig:fig11}\,(d) shows the total DOS at $p\,=\,$0, 1 and 2 GPa zoomed in around $E_F$. Under hydrostatic pressure, the unit cell volume is reduced, which increases the band overlap and hybridization. This means that the total energy range spanned by DOS increases, as also shown in Fig.\,\ref{fig:fig11}\,(d) by the shifting of the DOS peaks around $-$1 eV to lower energy and those around +1 eV to higher energy. As a result, the number of states at $E_F$, $N(E_F)$, is reduced with pressure. Another way to understand this is that the increased slope in band dispersions due stronger hybridization results in fewer states per energy. The total $N(E_F)$ is 8.8 state/eV/cell at $p\,=\,0$ GPa, and reduced to 8.5 at $p\,=\,$1 GPa and 8.2 at $p\,=\,$2 GPa. The calculated ambient-pressure value of $N(E_F)$ is consistent with the value of $N(E_F)\,=\,10.24\,$states/eV/cell, extracted from an experimental determination of the Sommerfeld coefficient via specific heat measurements \cite{Fujimori00}.
	
\section{Discussion}
	
	To gain further insight into the nature of superconductivity in NiBi$_3$, we check in the following whether our results of the pressure dependence of $T_c$ can be described in the framework of the Bardeen-Cooper-Schrieffer (BCS) theory of superconductivity. Within this theory, the critical temperature $T_c$ is determined by the Debye temperature $\Theta_D$, the electron-phonon coupling potential $V$ and the density of states at the Fermi energy $N(E_F)$ via
	
	\begin{equation}
	T_c = 1.136 \Theta_D \exp\left( -\frac{1}{V N(E_F)} \right).
	\label{eq:BCS}
	\end{equation}

Note that this approach is very simplified, as it presents a theory for weak-coupling superconductors and neglects other effects, such as electron-electron interactions. Even though there are more sophisticated approaches to model $T_c$ available, such as the McMillan equation, we restrict ourselves to the discussion of the simpler BCS equation due to the smaller number of unknown parameters.

At ambient pressure, $\Theta_D\,=\,144\,$K was determined from an analysis of specific heat data \cite{Fujimori00,Kumar11}. By using $N(E_F) (p\,=\,0\,$GPa) determined in the present work and the ambient-pressure $T_c$ of 4.03\,K, we infer an electron-phonon potential $V\,\approx\,30\,$meV. This value is of the order of $V$ values determined for other BCS superconductors \cite{Knoener17}.

\begin{figure}
	\centering
	\includegraphics[width=0.7\textwidth]{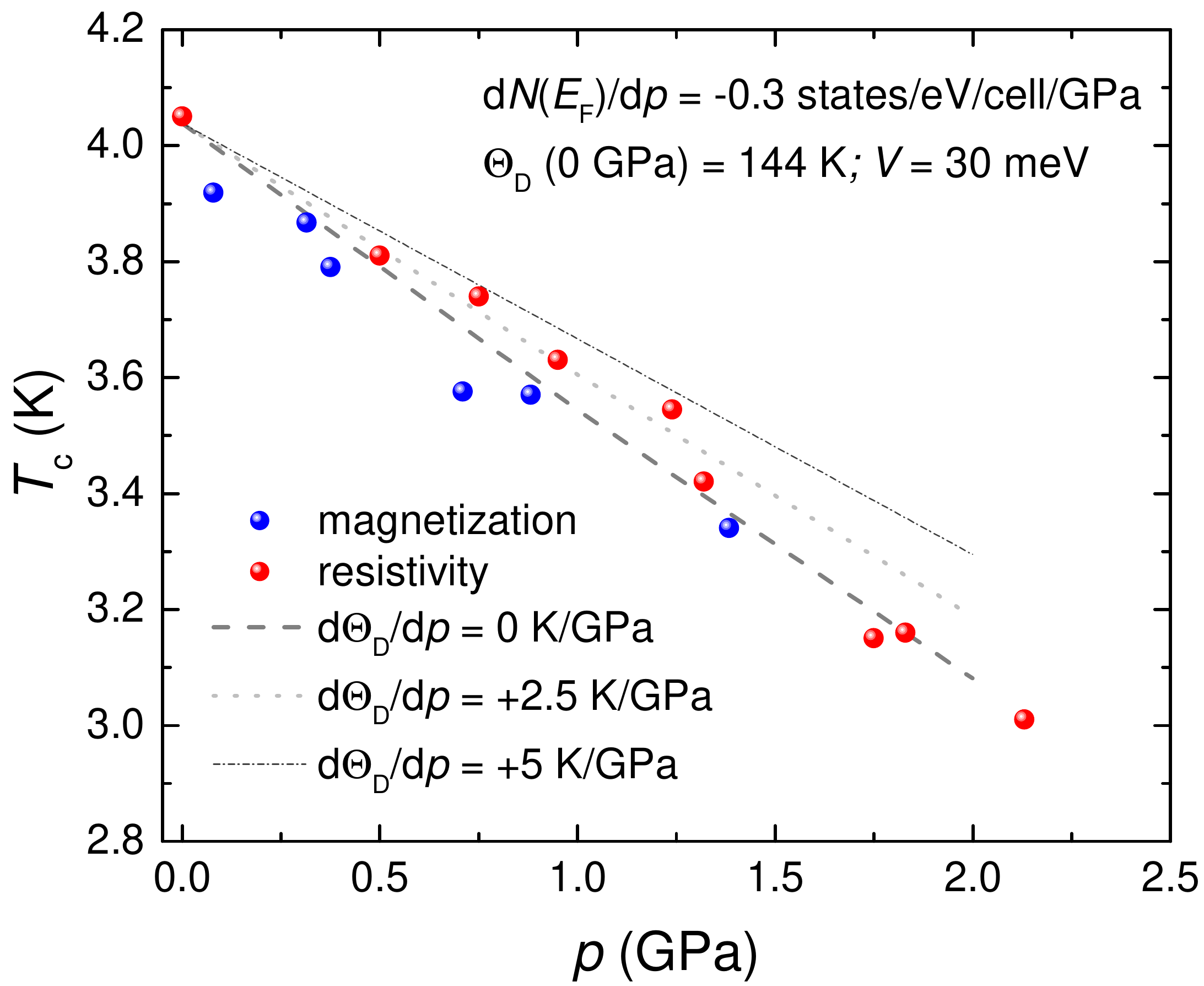} 
	\caption{Experimental data of $T_c$ vs. $p$, together with the fit of $T_c(p)$ by the BCS formula, given in eq. \ref{eq:BCS}. In order to obtain the fit, the theoretically-calculated change of the density of states $N(E_F)$ with pressure was used, as well as the Debye temperature $\Theta_D$ from literature \cite{Fujimori00}. The dashed line represents a fit for a constant $\Theta_D$ as a function of pressure, the dotted and the dashed-dotted line represent fits which take the pressure dependence of $\Theta_D$ into account (see main text for details).}
	\label{fig:fig12}
	\end{figure}

In order to evaluate the pressure dependence, we consider three cases: First, we consider the case in which the pressure dependence of $T_c$ is solely determined by the pressure dependence of $N(E_F)$. As outlined in Sec. \ref{sec:DFTresults}, the latter is given by d$N(E_F)$/d$p\,=\,-0.3$\,states/eV/cell/GPa. In the second and third case, we allow for a pressure dependence of $\Theta_D$. To this end, we assume that the pressure dependence of $\Theta_D$ is of the same order of magnitude as the pressure dependence of $T_{max}$. Therefore we chose two different pressure dependencies for our calculations (d$\Theta_D$/d$p$\,=\,+2.5\,K/GPa and d$\Theta_D$/d$p$\,=\,+5\,K/GPa). Note that d$\Theta_D$/d$p$\,=\,+5\,K/GPa most likely corresponds to the upper limit of the pressure dependence of $\Theta_D$. In all cases, we treat the electron-phonon potential $V$ as pressure-independent. The results of applying equation \ref{eq:BCS} for the three cases are shown as dashed, dotted and dashed-dotted lines in Fig. \ref{fig:fig12}, together with the experimentally-determined $T_c(p)$ data. We find a very good agreement of all three calculations to the measured experimental data. Importantly, all three scenarios yield a decrease of $T_c$ with $p$, thus demonstrating, that the decrease of density of states at the Fermi edge $N(E_F)$ with $p$ is responsible for the decrease in $T_c$. Thus, while also taking into account our observation of an almost pressure-independent Fermi velocity $v_F$, we summarize that superconductivity in NiBi$_3$ can be described in the framework of BCS theory. Therefore, NiBi$_3$ is most likely a conventional electron-phonon-mediated superconductor. Our conclusion is supported by a recent report of Andreev spectroscopy data \cite{Zhao18} which demonstrated singlet $s$-wave superconductivity in NiBi$_3$, as well as with analyses of the specific heat jump at $T_c$ \cite{Kumar11}.

\section{Summary}

In conclusion, we studied the properties of the superconductor NiBi$_3$ under pressure in a combined experimental and theoretical effort. Our results of resistivity and magnetization measurements under pressure indicate a moderate, slightly non-linear decrease of the superconducting critical temperature $T_c$ with pressure (d$T_c$/d$p\,\approx\,-$0.45\,K/GPa). The slight non-linearity of $T_c(p)$ is accompanied by a broad maximum in the normalized slope of the upper critical field, -(d($\mu_0 H_{c2}$)/d$T$)/$T_c$, indicating minor changes of the Fermi surface under pressure. The pressure dependence of $T_c$ can be described in the framework of BCS theory on a quantitative level indicating that the suppression of $T_c$ with $p$ can be attributed to a loss of density of states at the Fermi energy with $p$. Thus, we infer that NiBi$_3$ is a conventional electron-phonon mediated superconductor.

\section{Acknowledgements}

We thank Peter P. Orth for useful discussions. This work was carried out at Iowa State University and supported by Ames Laboratory, US DOE, under Contract No. DE-AC02-07CH11358. L. X. was supported, in part, by the W. M. Keck Foundation.

\bibliographystyle{unsrt}
\bibliography{Lit}

\end{document}